\documentclass[11pt]{article}
\usepackage{fullpage}
\usepackage{url}
\usepackage{amsmath,natbib}
\usepackage{graphicx,color}
\usepackage[T1]{fontenc}
\usepackage{lmodern}
\newcommand{\subdata}{Y_0}
\newcommand{\pridata}{Y}

\title{How to Incorporate Systematic Effects into Parameter Determination}

\author{David A. van Dyk\\ \normalsize
Statistics Section, Department of Mathematics, Imperial College London,   SW7 2AZ, UK\\
\tt{dvandyk@imperial.ac.uk}
\and 
Louis Lyons\\ \normalsize
Blackett Lab., Imperial College London, SW7 2BW, UK \\ \normalsize
Particle Physics, Oxford University, OX1 3RH, UK\\
\tt{l.lyons@physics.ox.ac.uk}
}

\begin{document}

\maketitle

%Some ideas for the title:
%\begin{itemize}
%    \item Accounting for Systematic Effects
%    \item Statistical Analyses in the Presence of Systematic Effects
%    \item A Short Review of Methods Accounting for Systematic Effects 
%\end{itemize}

\begin{abstract}

We describe two different approaches for incorporating systematics into analyses for parameter determination in the physical sciences. We refer to these as the Pragmatic and the Full methods, with the latter coming in two variants: Full Likelihood and Fully Bayesian. By the use of a simple and readily understood example, we point out the advantage of using the Full Likelihood and Fully Bayesian approaches; a more realistic example from Astrophysics is also presented. This could be relevant for data analyses in a wide range of scientific fields, for situations where systematic effects need to be incorporated in the analysis procedure. 

This note is an extension of part of the talk by van Dyk at the PHYSTAT-Systematics meeting\footnote{PHYSTAT-Systematics Workshop, November 2021, \url{https://indico.cern.ch/event/1051224/ .}}.

\end{abstract}

\section{Introduction}

This short note describes and compares two methods of incorporating the effect of systematics in analyses of data to extract parameter(s) of interest. 
In our setup, systematics take the form of the effect of ``nusiance'' parameters that are estimated from a subsidiary experiment or study on the primary experiment or study, where the parameters of interest are estimated. As a toy example, the nuisance parameter could be the intercept of a straight line and the parameter of interest could be the gradient. In a more realistic example, the nuisance parameter might be the sensitivity of an X-ray detector as a function of energy, and the parameter of interest might be the magnitude of and slope of a astrophysical X-ray energy spectrum. 
A ``na\"ive'' or ``default'' approach ignores the uncertainty in the estimates nuisance parameters when analyzing the primary experiment. In contract, the two methods we consider account for uncertainty in the estimated nuisance parameters, but differ in the way that they do so; we refer to the two methods as the Pragmatic and the Full Likelihood (or Full Bayesian) approaches. 

The first section of this note introduces the concept of systematics, while Section~2 lists some of the methods used for incorporating systematics in data analyses. The straight line fit is explored in some detail in Section~3. This is followed by an Astrophysics example in Section~4. Finally some conclusions appear in Section~5.

\subsection{What are systematics?}
Almost every measurement of a physical quantity is subject to two sorts of uncertainties: statistical and systematic.

Random or statistical uncertainties result from the limited accuracy of measurements, or from the fluctuations
that arise in counting experiments where the Poisson distribution is relevant. If the experiment
is repeated, the results will vary somewhat, and the spread of the results in such replicates provides an estimate
(but not necessarily the best)  of the statistical uncertainty. 

%{\bf\color{blue} Louis can you incorporate the "subsidery" and "primary" experiment into this:} 
Systematic uncertainties can also arise in the measuring process. The quantities we measure may be shifted from the true values.
For example, our measuring device may be miscalibrated, or the number of observations we count may be not only from the 
desired signal, but also from various contamination sources.  Such effects would bias our result, and we should correct for them, for
example by performing some calibration measurement; the systematic uncertainty arises from the remaining uncertainty in our corrections. We refer to the calibration as the `subsidiary measurement' as compared with the `main experiment' for the parameter of interest. 

Systematics can cause a similar shift in the result for a repeated series of experiments, and so, in contrast to statistical uncertainties, they may not be detectable by looking for a spread in 
the results, and do not necessarily shrink in magnitude as more data are collected.

\subsection{Systematics for a pendulum experiment}
As a simple example, consider a (primary) pendulum experiment designed
to measure the acceleration due to gravity $g$ at sea level in a given location,
\begin{equation}
g = 4\pi^2 L/\tau^2,
\label{Pendulum}
\end{equation}
where $L$ is the measured length of the pendulum, $\tau = T/N$ is its period, and $T$ is the measured time for $N$ oscillations.
The statistical uncertainties are the  ones affecting the measurements of $L$ and 
$T$\footnote{Although $N$ involves counting the number of swings, we 
do not have to allow for Poisson fluctuations, since there are no random 
fluctuations involved.}.

There may also be systematic uncertainties on these variables. These can be estimated by performing subsidiary measurements to calibrate our ruler and clock.

Unfortunately there are further possible systematics not associated with the measured 
quantities, and which thus require more careful consideration. For example, the 
derivation 
of eqn. (\ref{Pendulum}) assumes that 
\begin{itemize}
\item{our pendulum is simple i.e.  the string is massless, and has a massive bob of infinitesimal 
size;}
\item{its support is rigid;}
\item{the oscillations are of very small amplitude $\theta$ (so that 
$\sin\theta \approx \theta)$; and}
\item{they are undamped.}
\end{itemize}
None of these are exact in practice, and so corrections must be estimated for 
them. The uncertainties in these corrections are systematics.

Furthermore, there  may be theoretical uncertainties. For example, we may want the 
value of $g$ at sea level, when the measurements are performed on top of a mountain. We thus 
need to apply a correction, which depends on our elevation and on the local geology.
There might be two or more different estimates of this theoretical correction factor, and again this source of error/uncertainty  
contributes a systematic uncertainty.
\vspace{1.0ex}

Realistic analyses are in general more complicated (or extremely more complicated) than this,
and often require the construction of a likelihood function and/or using some fitting procedure 
to extract the parameter(s) of interest. From here on, we refer to the parameters of interest as $\phi$, the nuisance parameters (leading to systematic effects) as $\nu$, the data from the subsidiary experiment as $\subdata$ and the data from the primary experiment as $\pridata$.

\section{Methods of incorporating systematics in data analyses}
In order to incorporate a systematic effect into our analyses, three activities are required. 
%\dvd{Is this something other than identifying and estimating the nuisance parameters?}
First, we have to identify what possible sources of systematic effects may be relevant. Then we need to estimate the magnitude of the systematic effect. In general, these vary with the type of analyses involved, and requires detailed knowledge of the experimental procedure. They are not considered further here. 

Finally, we have to incorporate this numerical information into the procedure for estimating the parameter of interest, $\phi$. This was the topic of the PHYSTAT-Systematic Workshop$^1$ %\footnote{PHYSTAT-Systematics Workshop, November 2021, \url{https://indico.cern.ch/event/1051224/ .}} 
where possible methods such as those below were discussed in detail.

%\dvd{Can we also mention the subsidiary experiment  and the main experiment, as well as the data from both -- or the data just from the main experiment?}

\subsection{List of Methods}
In this section, we merely list and explain very briefly some possible methods for including systematics in an analysis. Further details are available from the PHYSTAT Workshop 
\begin{itemize}

\item{Simple uncertainty propagation, with One Parameter At a Time (OPAT):

There are two variants of this method. 
In the first, the nuisance parameter $\nu$ is moved from its central value by its uncertainty $\sigma_{\nu}$, and the change in the result
is used as the contribution to the systematics. 
If the magnitudes of the changes in the result for upward and downward shifts of $\nu$ differ, the assumption of the result varying linearly with $\nu$ is inadequate. Also there are various prescriptions for deciding what the contribution from a particular systematic should be when a change in result is estimated as $\Delta\phi \pm u$, where $u$ is the uncertainty on the shift $\Delta\phi$ \citep{HL}.   

Alternatively, $\nu$ is sampled from its expected distribution many times,
and for each of these the analysis procedure is repeated; the width of the distribution of results is the contribution to the systematics. 
%\dvd{Can we add a citation for this method?}

If there are several independent sources of systematics, with the contributions from each being $\sigma_i$, then the overall
estimated systematic $\sigma_t$ is given by
\begin{equation}
\sigma_t^2 = \sum_i \sigma_i^2
\end{equation}  
This assumes that the dependence of the result on the nuisance parameters is additive. If not, the procedure of the next bullet is preferable. }

\item{Simple uncertainty propagation, with several nuisance parameters varied together:}

Here all $n_{\nu}$ parameters are varied simultaneously. In analogy with the OPAT example above, this can be achieved 
by selecting specific points at grid values in the hypervolume of the $n_{\nu}$-dimensional space; or by taking a large number of 
random samples throughout the hyperspace. 
 
\item{$\chi^2$:

The weighted sum of squares is defined as 
\begin{equation}
S = \sum_i\sum_j (\pridata_i - m_i)C^{-1}_{ij}(\pridata_j-m_j),
\end{equation} 
where here $\pridata$ are the data from the main and the subsidiary measurements, 
%\dvd{from the primary experiment?} 
$m$ is the model prediction of the data and
$C$ is the covariance matrix, including systematic variances and correlations.

This is equivalent to including Gaussian constraint terms in the likelihood, corresponding to the information from any subsidiary measurements for the nuisance parameters which affect the $m_i$.}

\item{Likelihood:

Here, the effect of a nuisance parameter $\nu$ is incorporated in the likelihood as an additional factor which constrains $\nu$ to be close to its measurement in a subsidiary experiment. For example, in a counting experiment the number of expected events may depend not only on the assumed signal strength $\phi$ but also on an uncertainty from background effects.
%\dvd{what is $\nu$ in this example?}  
Then  with $\pridata$ observed events in the primary experiment, the likelihood is given by

\begin{equation}
%{\mathcal L}(\phi,\nu) = P(\pridata;\phi,\nu)*G(\nu _{\rm meas};\nu)
{\mathcal L}(\phi,\nu) = P(\pridata;\phi,\nu)*G(\hat\nu(\subdata);\nu)
\end{equation}
where $P$ is the Poisson probability of observing $\pridata$ events when the parameters have values $\mu$ and $\nu$; and $G$ is a Gaussian constraint for obtaining an estimate of $\hat\nu$ based on the subsidiary measurement 
%$\nu _{\rm meas}$ 
$\subdata$
when the true value is $\nu$.
Finally the profile likelihood is used to obtain an interval or a limit on the parameter of interest $\phi$. %\dvd{Is there a citation for using profile likelihood in this context? Say how the uncertainty quantified?} 
Here 
\begin{equation}
{\mathcal L} _{\rm prof}(\phi) = {\mathcal L}(\phi,\nu _{\rm best}(\phi))
\end{equation}
where $\nu _{\rm best}(\phi)$ 
%\dvd{This is not the notation used in Eq (5). Change (5) or this?} 
is the value of $\nu$ which maximises the ${\mathcal L}$ for each value of $\phi \ $ i.e. $\nu _{\rm best}(\phi)$ is a function just of $\phi$; and the profile likelihood ${\mathcal L} _{\rm prof}(\phi)$ is a function of only $\phi$, and not of $\nu$.  }

\item{Bayes:

The standard Bayesian approach is to multiply the likelihood for the parameter of interest $\phi$ by its prior probability distribution $\pi(\phi)$; this product is proportional to the posterior probability distribution $p(\phi \mid \pridata)$. 

With a nuisance parameter $\nu$, the likelihood ${\mathcal L}(\phi, \nu)$ is now a function of both $\phi$ and $\nu$. As before, we assume that information about $\nu$ is derived from a subsidiary measurement; application of Bayes' Theorem yields
\begin{equation}
p _{\rm sub}(\nu) = {\mathcal L} (\subdata;\phi, \nu) * \pi(\nu)
\end{equation}

%\dvd{These two equations don't make sense to me. if you substitute (6) into (7), for example, the likelihood appears twice! I think we need two Likelihood functions -- one from the primary analysis and one from the subsidiary. OR just rely on $p _{\rm sub}(\nu)$, without deriving it.}
where $\pi(\nu)$ is the prior for $\nu$, and $p _{\rm sub}(\nu)$ is its posterior from the subsidiary experiment. It is this that serves as the prior 
%$\pi(\nu)$ 
for the main the experiment i.e.
\begin{equation}
p(\phi,\nu) = {\mathcal L}_{main}(\pridata;\phi,\nu) * \pi(\phi) * p _{\rm sub}(\nu) 
\end{equation}
Then the posterior for the parameter of interest $\phi$
can be obtained by integrating the joint posterior distribution $p(\phi, \nu)$ over $\nu$. From this, the Bayesian upper limit or credible intervals can be obtained at any desired credible level.}

\item{Frequentist Neyman construction: 

The Neyman construction \citep{Neyman} is used to obtain a frequentist confidence interval for the parameter(s) of interest $\phi$, given the observed data. It automatically ensures that the intervals have the correct coverage. That is, in a series of repeated repetitions of the measurement, it is guaranteed that a specified fraction of the set of intervals will contain the true value(s) of $\phi$.

To make the procedure unique, an ordering rule is required for the Neyman construction. In this way, different types of intervals can be produced, e.g., Upper Limits, Lower Limits, or Central Intervals. A common rule in Particle Physics uses likelihood-ratio ordering \citep{FC}, which produces `Unified Confidence Intervals'.

A disadvantage of  Neyman intervals is that as the number of parameters of interest plus nuisance parameters increases, the computational effort soon becomes intractible. A recent paper by the NOvA Collaboration \citep{NOvA} gives details of how this can be dealt with, while maintaining approximate coverage.}

\item{Mixed Frequentist/Bayes:

This is a procedure where the parameter of interest is treated in a frequentist way, but a Bayesian approach is used for the nuisance parameters. The paper by Cousins and Highland \citep{CH} describes how to do this for upper limits.} 
\item{Pivots:

The idea is to construct a variable involving the systematic such that the analysis turns out to be independent of or insensitive to the value of the systematic, while it still maintains power for discriminating among different values of the parameter of interest.  }

\item{Machine Learning techniques:

These have recently made big inroads into many aspects of Particle Physics Analyses. For their several usages in dealing with systematics, see the video and slides of Kyle Cranmer's talk at PHYSTAT-Systematics.}

\end{itemize}

%For methods that determine effect of
%independent sources of systematics,
%total syst=$\sqrt($Sum of squares of independent effects)

%But using full likelihood, as product of likelihoods from main and  subsid expts, is better
%*************** CONTINUE HERE ********

In this note, we compare simple uncertainty propagation, which we refer to as the  Pragmatic approach, with either the  Full Likelihood method (see Sections 3.2 and 3.3) or with the Full Bayesian treatment, as in the Astrophysics example of Section~4.

%\dvd{Can we now say something about the Naive, Pragmatic, and Fully Bayesian approaches? That is, define them and compare them to the methods mentioned above? This could be similar to the three points on page 12.}

%***********************THE TEXT BELOW IS AN OLDER ALTERNATIVE TO THE TEXT BELOW
%
%In order to incorporate a systematic effect into our analysis, two activities are involved:
%\begin{itemize}
%\item{Estimate the magnitude of the systematic effect. In general, this varies very much with the type of analysis involved, and %requires detailed knowledge of the experimental procedure.} 
%It is not considered further here. }
%\item{Incorporate this numerical value into the procedure for extracting the parameter of interest $\phi$. This was the topic of the %PHYSTAT-Systematic Workshop \citep{PHYSTAT-S}, where possible methods were discussed in detail.   } 
%\end{itemize} 

\section{A simple example: Straight line fitting}
\label{sec:line}
We now employ a very simple example to demonstrate the differences between the Pragmatic and the Full Likelihood approaches for dealing with systematics.
It consists of fitting a straight line $y =  a +b*x$ to a series of data points, which provide measured values $y_i \pm \sigma_i$ at a series of specified $x$ values  ($x_i$) $-$ see the red points in Fig.~1. (For Particle Physicists this could be thought of as fitting a straight line 2-dimensional track, to a series of measurements in tracking detectors at precisely known locations $x$). For simplicity, we assume that all the $\sigma_i$ are equal. We are interested in the value of the gradient, so $b$ is our
parameter of interest  $\phi$, while the intercept $a$ is merely a nuisance parameter $\nu$. The problem is that the measured points are close together in $x$, which makes it hard to determine the gradient. If we consider possible lines of slightly different gradients, they would have different intercepts; the parameters $a$ and $b$ in our example are strongly correlated. With the mean of the $x_i$ values being positive, this correlation is negative (i.e. increasing $b$ results in a lower value of $a$).

%Fig 1: Tracking example withred and 1 blue point
%\clearpagehttps://www.overleaf.com/project/63e658a7eb3a7263cee341b2
\begin{figure}[ht]
\begin{center}
\includegraphics[width=0.8\textwidth]{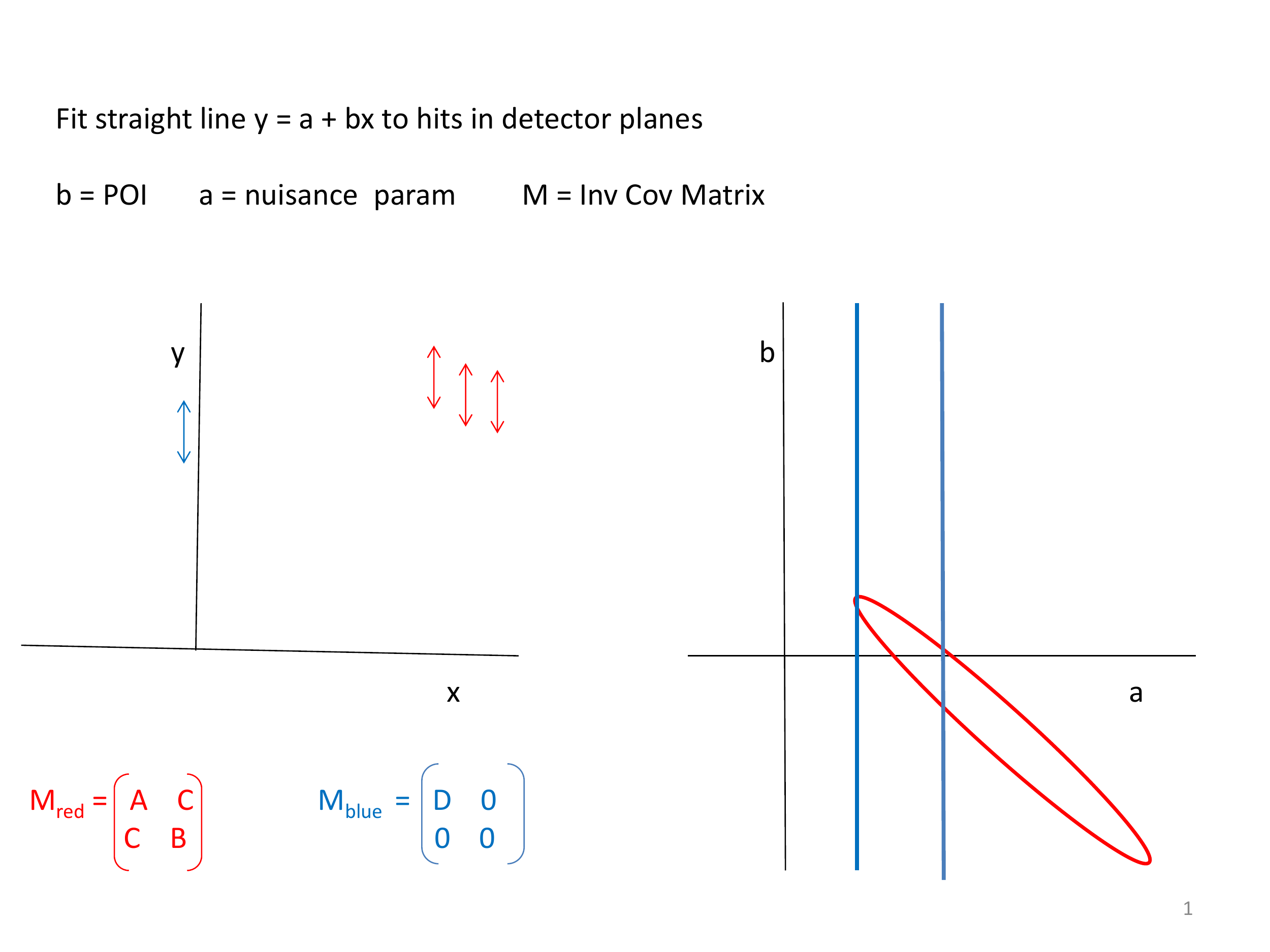}
\caption{Tracking example: The left diagram shows the 3 measurements of the main experiment (in red) and the single measurement of the subsidiary one (in blue). It is clear that the gradient of a straight line fit just to the red points has a large uncertainty, while combining both sets of points results in a much better determination. Note also that the fit to just the red points yields a negative gradient, while the fit to all 4 points has a positive one.
The right-hand plot shows the covariance ellipse for the main experiment in red. For the subsidiary measurement, the vertical axis
 of the `ellipse' is infinite in length, as there is no information about the gradient $b$ of the line  from the single  blue point; its
preferred region in parameter space $(a,b)$ is between the two vertical blue lines.}
\label{DvD_1}
\end{center}
\end{figure}

 One way of improving our knowledge of $b$ is to perform a subsidiary measurement to provide  an independent estimate of the nuisance parameter $a$. This is provided by the blue point in Fig 1.  It is clear intuitively why this helps: whereas the three red points of our main experiment were closely spaced, the 4 points cover a wider range of $x$, and thus provide a much improved estimate of the gradient.

We also see that in our example the value of the best fit gradient to the red points is very different from that of the better fit to the 4 points. A problem with the Pragmatic approach is that it does not update the gradient from the main experiment with the new information from  the subsidiary one. (This is especially so if there are more than one sources of systematics.)
\vspace{1.0ex}

We next examine the estimated systematic uncertainties more formally from the viewpoint of the Pragmatic approach and the Full Likelihood one.

\subsection{Pragmatic approach}
Here the gradient is determined from the main experiment, i.e. just from the three red points of Fig 1. Its statistical uncertainty is obtained by assuming some fixed value for the nuisance parameter $a$.  Information about $a$ comes directly from the subsidiary experiment, with uncertainty $\sigma_{a, {\rm blue}}$. This then is propagated to  become the systematic uncertainty on $b$, by seeing how much the best value of the gradient $b$ changes when $a$ is changed  by $\sigma_{a, {\rm blue}}$. 

In general, the Pragmatic  approach could be used for more than one systematic effect.

The best value of the parameter of interest $\phi$ and its statistical uncertainty $\sigma _{\rm stat}$ 
are taken from  the main experiment. For the specific example of straight line fitting, this corresponds to the red points 
and the red ellipse of Fig.~\ref{DvD_1}, with $\phi$ being the gradient $b$.  At this stage, the potential nuisance parameters $\nu$ 
(i.e. the intercept $a$ of the straight line) are kept fixed, presumably at some current estimate 
of their values. 
%i.e. the uncertainties in the $\nu$ are at this stage not taken into account. 
Then the effect of each potential nuisance parameter on $\phi$ is estimated; the change in $\phi$ as a particular nuisance parameter $\nu_i$ is changed by its uncertainty is taken as the contribution to the systematic uncertainty $\sigma_{{\rm syst},i}$ from that particular source. 

Finally the square of the total systematic uncertainty $\sigma _{\rm syst}^2$ from several uncorrelated sources of systematics is taken as the sum the squares of these individual contributions\footnote{If the different sources are correlated, then 
$\sigma _{\rm syst}^2 = \Sigma\Sigma\sigma_{{\rm syst},i} M_{ij} \sigma_{{\rm syst},j}$,
where $M$ is the inverse covariance matrix for the different sources of systematics.}.    %As usual, the square of the 
%total uncertainty is the sum of the squares of the statistical and the systematic uncertainties.

\subsection{Full Likelihood approach}

This consists in first writing down the probability density distribution
\begin{equation}
p(y;a,b,x) = p _{\rm main}(y_{\rm main};a,b,x _{\rm main}) * p _{\rm subsid}(y_{\rm subsid};a,b,x _{\rm subsid})  
\label{pdf}
\end{equation}
for observed data values of $y$ at the specific $x_i$ of our straight line example, for both the main and the subsidiary measurements (i.e. for the red and the blue points respectively of Fig.~\ref{DvD_1}) , assuming fixed values of the parameters $b$ and $a$. We assume that the individual $y$ are Gaussian distributed around their predicted values $y _{\rm pred} = a + b*x_i$  with variance $\sigma_i^2$:
\begin{equation}
p(y_i;a,b,x_i) = 1/(\sqrt(2\pi)\sigma_i) * exp[-0.5(y_i -  (a+b*x_i))^2]/\sigma_i^2].
\end{equation}

This is then turned into the likelihood function:
\begin{equation}
\label{Likelihood}
{\mathcal L}(a,b;y) =\Pi p(y_i; a,b,x_i)
\end{equation}
where the product is over all the observations in the main and subsidiary experiments. Thus in eqn. \ref{pdf} for $p$, 
the parameters are taken as being fixed and it provides the probability density of different data values $y$; 
the likelihood of eqn. \ref{Likelihood}  has the data as fixed and is regarded as a function of the parameters.

It is  perhaps not surprising that in this case the parameters  determined from the full likelihood approach agree with those obtained by regarding the main and subsidiary experiments as independent determinations of  the parameters, and then combining these two pairs of values by %BLUE \citep{BLUE} or by 
a $\chi^2$ procedure - see Section~\ref{Numerics}.

\subsection{Detailed comparison}
\label{Numerics}
\subsubsection{Algebra}
Here we derive formulae for the uncertainty on $b$, as determined in the two approaches. This is most easily achieved by using 
the inverse covariance matrix {\bf M} =
$\begin{pmatrix}
A & C \\
C & B
\end {pmatrix}$.
The log-likelihood function -2*ln ${\mathcal L}(a,b)$ for our fit to the main experiment will have contours in the plane of our parameters $a$ and  $b$ such that
\begin{equation}
\label{ellipse}
Aa^2 + Bb^2 + 2Cab = {\rm constant}  
\end{equation}
This is the equation of an ellipse. With the constant set equal to unity, this ellipse is suitable for determining the $68\%$ confidence level  uncertainties on each of the parameters separately; they are given by half the total width of the ellipse in the $a$ or in the 
$b$ directions. For the area inside an ellipse to contain the true values of $a$ and $b$ at that confidence
level, the constant needs to be set at 2.3.

If we invert the inverse covariance matrix, we obtain the covariance matrix, whose diagonal elements are the variances of $a$ 
and of $b$, and the off-diagonal one is their covariance. (See, for example, ref.  \citep{LL_book}  for a longer discussion of covariance matrices.) 

% For the primary experiment, the inverse covariance matrix is denoted as
%$\begin{pmatrix}
%A & C \\
%C & B
%\end {pmatrix}$.
 For the straight line fit of the primary experiment, 
\begin{equation}
\label{Mred}
A = \Sigma (1/\sigma_i^2 ),\ \ \ \ \ \ \  B = \Sigma (x_i^2/\sigma_1^2),\ \ \ \ \ \ \  C = \Sigma (x_i/\sigma_i^2)
\end{equation}
where the summations are over the red data points.
For the secondary experiment, the matrix is
$\begin{pmatrix}
D & 0 \\
0 & 0
\end {pmatrix}$.
Here, $D$ is given by the same as the expression for $A$ in the equation above, but just for the single blue point. 

Some of the properties of covariance matrices are shown in Figs 2 and 3.  
In particular the full uncertainty on $b$ is the half-height of the smallest rectangle with sides parallel to the $a$ and $b$ axes, and enclosing the ellipse. The statistical uncertainty is given by half the distance between the values of $b$ on the ellipse, when the  nuisance parameter is kept fixed at its best value.  

In Fig.~\ref{DvD_3}, two straight lines are drawn through the centre of the ellipse. We refer to these as the `Profile' line and the `Minimisation' line. The former is used for evaluating the profile likelihood; this is the value of the likelihood as a function of
the parameter of interest  $b$ when, at each value of $b$, the nuisance parameter $a$ is varied to provide 
the best value of the likelihood. It is used for determining the total uncertainty on $b$. Its gradient on the $b$ versus $a$ 
plot  is $-A/C$.  
In contrast, the `Minimisation' line shows how $b$ changes when the nuisance parameter $a$ varies. It has gradient $-C/B$. It is used  in the calculation of the systematic uncertainty on $b$ in the Pragmatic approach, arising from the uncertainty in $a$ from the subsidiary measurement.

%%
%Fig 2: Properties of Covariance ellipses
\begin{figure}[ht]
\begin{center}
\includegraphics[width=0.8\textwidth]{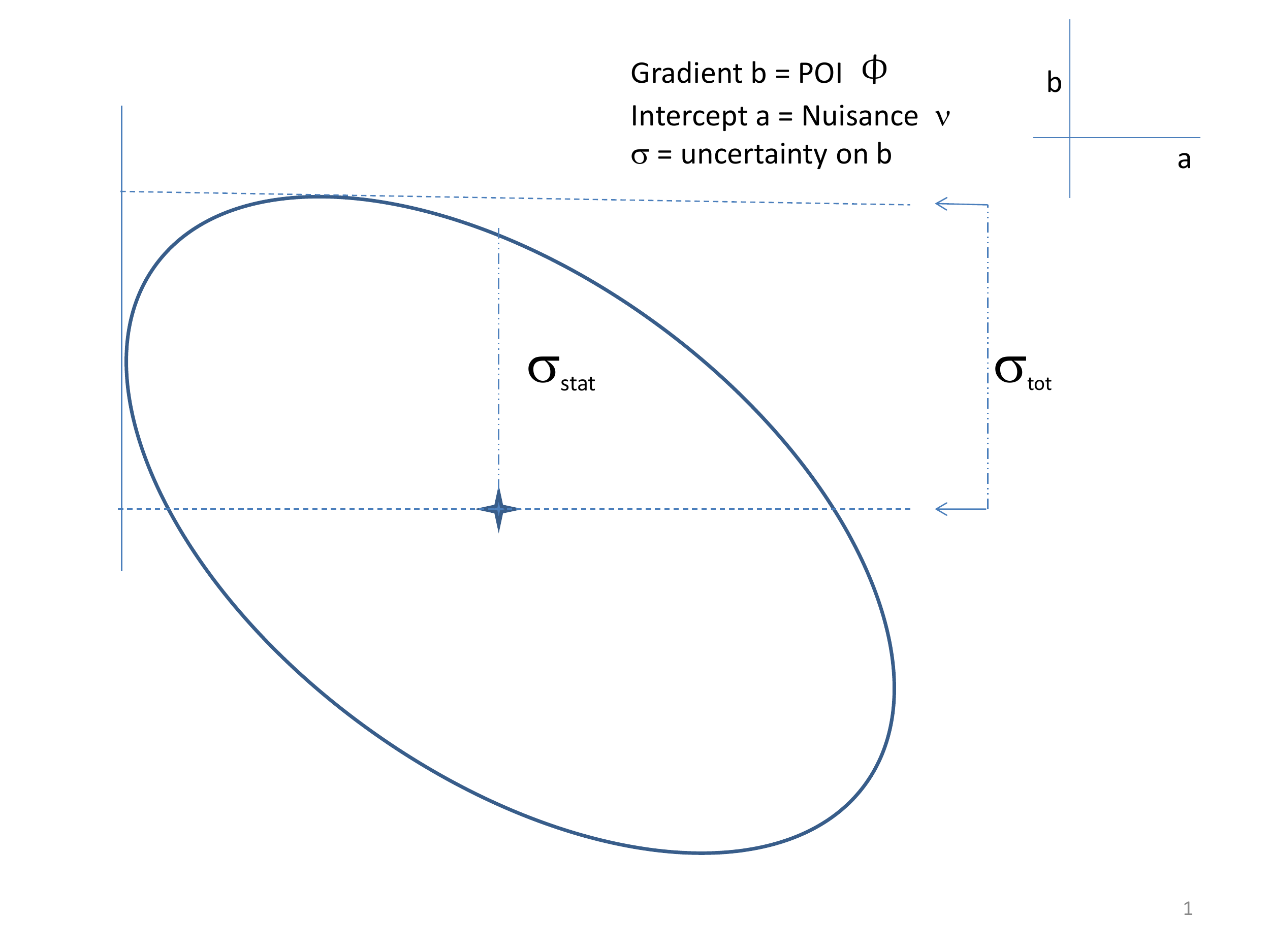}
\caption{Properties of Covariance Ellipses: 
The contour of $\Delta\ln{\mathcal L} = 0.5$ in $(a,b)$ space. The statistical uncertainty for $b$ ($\sigma _{\rm stat}$) is obtained
assuming that the nuisance parameter $a$ is known exactly, and is $1/\sqrt B$.  The total uncertainty in each variable is given by half the height or width of a rectangle with sides parallel to the axes, and which just encloses the ellipse.
For the Full Likelihood case, the contour corresponds to the combined likelihood for the main and subsidiary measurements.}  

\label{DvD_2}
\end{center}
\end{figure}

\vspace{0.1in}

%%
%Fig 3: Pragmatic approach
\begin{figure}[ht]
\begin{center}
\includegraphics[width=0.8\textwidth]{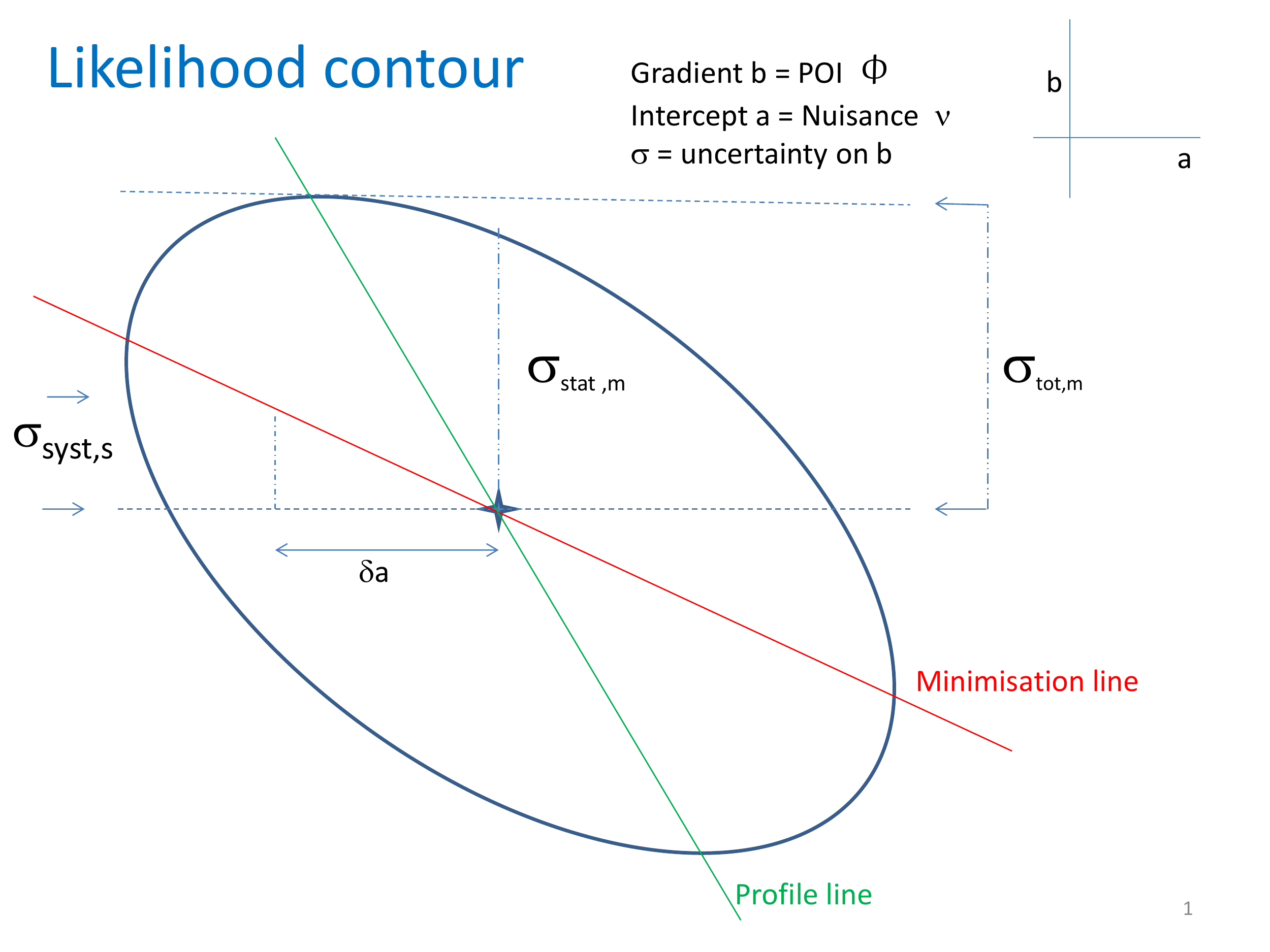}
\caption{In the Pragmatic approach, the likelihood contour corresponds just to the main experiment. The statistical uncertainty on $b$ is $\sigma_{stat,m}$, and the total uncertainty just from the main experiment is $\sigma_{tot,m}$. As a result of the subsidiary experiment, the systematic  uncertainty for $b$ 
($\sigma _{\rm syst,s}$) is given by the change in $b$ when $a$ is varied by its uncertainty from the subsidiary experiment i.e. by $\delta a =
1/\sqrt D$; this is not connected with the covariance ellipse of the main experiment. The systematic shift in $b$ also 
depends on the gradient of the `Minimisation' line, which is $-C/B$.  The systematic is thus estimated as $C/(B\sqrt D)$. }
\label{DvD_3}
\end{center}
\end{figure}

For the Pragmatic approach,
\begin{equation}
\label{Prag}
\sigma _{\rm stat}^2 =1/B,\ \ \ \ \ \ \ \ \sigma _{\rm syst}^2 = C^2/(B^2D), \ \ \ \ \ \  \sigma _{\rm tot}^2 =  (1/B)*[1 + C^2/(BD)]
\end{equation}
These are obtained by considering
two different fits of the straight line which are performed to the 3 data points of the main experiment, first with a fixed value $a_0$ for the intercept, and then with it changed to $a_0 + \sigma_a$, where $\sigma_a$ is the uncertainty on $a$ as measured in the secondary experiment. In each of these fits, there is only one free parameter $b$.
Then the uncertainty on the result of the first fit
is taken as the statistical uncertainty on $b$, while the difference in $b$ between the two fits is its systematic uncertainty. 
%Interestingly, this produces the same results as in eqn. \ref{Prag}.

In contrast with eqn. \ref{Prag}, with the Full  Likelihood
%\begin{equation}
\begin{align}
\sigma _{\rm stat}^2 &=1/B,\ \ \ \ \ \ \ \ \ \ \sigma _{\rm syst}^2 = \sigma _{\rm tot}^2 - \sigma _{\rm stat}^2, \\
\sigma _{\rm tot}^2 &=  (A+D)/(B(A+D) -  C^2)  = (1/B) *[1 -C^2/(B(A+D))]^{-1}
\label{Full}
\end{align}
%\end{equation}
The result for $\sigma _{\rm tot}$ comes simply from inverting the inverse covariance matrix 
$\begin{pmatrix}
A+D & C \\
C & B
\end {pmatrix}$
for the combination of the main and the subsidiary measurements.

\vspace{0.1in}
It is interesting to compare $\sigma _{\rm tot}$ for the two approaches in eqns. \ref{Prag} and \ref{Full}. In the situation where the terms in the square brackets are not much larger than unity, the $D$ in eqn. \ref{Prag} is replaced by $A+D$ in eqn \ref{Full}. This has the effect of making the systematics uncertainty larger for the Pragmatic approach. As $D$ becomes small (corresponding to a very large uncertainty in the subsidiary measurement), the systematic uncertainty in the Pragmatic method tend to infinity, while for the Full Likelihood it tends to a constant. 

%******** Put in some numerical values to illustrate the importance or otherwise of the difference.

%******* DvD: Sometimes I consider just one variable ($b$) and at others two variables ($b$ and $a$). Do I need to worry about different values of $\Delta(lnL)$?   

\subsubsection{Numerical example}
Here we use a specific experimental example, to give numerical values of the systematic uncertainties, to illustrate how the two approaches compare. We use the situation shown on the left side of Fig 1, with the following parameters:
\begin{itemize}
\item{The main measurement has 3 detector planes at $x=3,\ \ x=4$ and $x=5.$ The uncertainties on the y-values are all 1.0. }
\item{The subsidiary measurement has a single detector plane at $x=0$. The uncertainty on its $y$ measurement is varied between 0.25 and 4.0.}
\end{itemize}
Our parameter of interest is the gradient $b$ of the straight line fit to the measurements, with systematic effects arising from the intercept $a$ of the fit. The values of the squares of the systematic uncertainties on $b$ are shown in Table 1.

%\vspace{0.2in}
\begin{table} [h!]
\begin{center}
\caption{The table compares the squares of the systematic uncertainties $\sigma _{\rm syst}^2$ on the gradient $b$ of the straight line, for the Pragmatic and Full Likelihood approaches. They refer to the situation  shown in the left side of Fig 1, and as described in the text. The column labelled $\sigma_0$ shows the value assumed for the statistical uncertainty on $y$ for the 
single measured point in the subsidiary experiment; the corresponding value for all the points of the main experiment is unity.
In each case, the square of the purely statistical uncertainty is 0.0130.} 
\begin{tabular}{rcll}
\hline\hline
$\sigma_0$  &&  $(\sigma _{\rm syst}^{\rm Prag})^2$   &  $(\sigma _{\rm syst}^{\rm Full})^2$ \\[3pt]
\hline
%0.25    & 0.00237  &  0.00236 \\
%0.5     & 0.0095      &  0.0093    \\
%1.0     & 0.038      & 0.35     \\
%2.0     &0.052       & 0.116    \\
%4.0     &  0.61    &  0.27   \\
0.25    &&  0.034  & 0.034 \\
0.5      &&  0.049  & 0.047   \\
1.0      &&  0.078  & 0.071   \\
2.0     &&  0.14     & 0.11   \\
4.0      && 0.25     & 0.18    \\
8.0    &&   0.48    & 0.26    \\
16.0   &&   0.94   & 0.34    \\
\hline\hline
\end{tabular}
\end{center}

\end{table}

The main points to notice are:
\begin{itemize}
\item{The Pragmatic systematic uncertainties are all larger than those of the Full Likelihood approach. They do, however become similar as $\sigma_0$ decreases.}
\item{As $\sigma_0$ increases, not surprisingly the systematic uncertainty for both methods increase. However, that for the Pragmatic approach becomes even larger than the systematic estimate just using the data from the main experiment 
($(\sigma _{\rm syst}^{\rm main})^2 = 0.487$). This is a very undesirable feature, which the Full Likelihood method avoids. The Pragmatic approach should not be used in such situations.}
\end{itemize}

\subsubsection{Other considerations}
\label{Other}
The Pragmatic method is usually simpler to apply in practice, especially for situations where the systematic is complicated. As an example from Particle Physics, most analyses using data from high energy accelerators  need detailed information about the behaviour of quarks and gluons within a proton; these are parametrised in the so-called `Parton Distribution Functions'. They are derived from global analyses of many other high energy processes, and also requires significant theoretical input; a whole industry is devoted to this endeavour. The uncertainties of such procedures are a source of systematic for the main analysis. The Pragmatic approach greatly simplifies the way in which these are incorporated in analyses.

Another reason for possibly using the Pragmatic approach is provided in the literature on Bayesian modeling, which recognizes that
potential misspecification of parts of a multicomponent Bayesian
model can justify ``cutting feedback'' or ``modularization'',  i.e.
informally limiting the influence of some model components
(see, for example, \citep{Jacob} and references therein).
Here this corresponds to not fully using the information from the subsidiary
measurement to update the central value of the parameter of interest
from the main measurement.  This results in larger systematic
uncertainties, reflecting concern that the impact of information
from the subsidiary measurement is not described with complete fidelity
by the full model.

%Also, Tom Loredo\footnote{Remarks made at the 2023 Banff Workshop on `Systematics' (23w5096).} has pointed out that there can be advantages from the Pragmatic approach not using the information from the subsidiary measurement to update the central value of the parameter of interest from the main measurement, and also returning a larger estimate of the systematics. If the information from the subsidiary measurement is regarded as less reliable than that from the main one, this might be a sensible conservative option.  

%\section{Example from Astrophysics}
%**************DvD to write the Astrophysics example from PHYSTAT-Systematics Workshop%

\section{An example from Astrophysics}
\label{sec:astro}

We now discuss a more realistic example from Astrophysics, involving a simplified simulation of results from the Chandra X-ray Observatory\footnote{\url{https://chandra.harvard.edu}}. It analyses an X-ray energy spectrum, where the instrument records photon counts in each of 1078 energy bins. (For simplicity we ignore the effect of errors in the recorded energy of photons, i.e., the photon redistribution matrix; a more extensive version of this simulation can be found in \citet{xu:etal}) The sensitivity of the bins vary according to their `effective area'. The simulation assumes a power law source spectrum so that 
\begin{equation}
{\mathcal E}(Y_i) = a_i T \alpha  (E_i/E_0)^{-\beta}
%{\rm for i = 1, ‚Ä¶ 1078      ********}
\label{dvd.model}
\end{equation}
where $i = 1, \ldots, 1078$ indexes the energy bins; 
${\mathcal E}(Y_i)$ is the expectation of the photon count, $Y_i$;  
$E_i,\ a_i,$ and $Y_i$ are the energy, effective area  (including efficiency), and photon count for each bin $i; 
\ \alpha$ and $\beta$ are the powerlaw parameters of scientific interest; 
$T$ is the exposure time;  
and the vector $A = (a_1, \ldots, a_{1078})$ is only known approximately, i.e., with error.  
$E_0$ is an arbitrary energy scale; its value does not affect the value of $\beta$, but it does affect that of $\alpha$, and also the magnitude and even the sign of the correlation between $\alpha$ and $\beta$. 
We assume that the observed photon counts are distributed as independent Poison variables with expectation given in Equation~\ref{dvd.model}; we denote the collection of counts by $Y=(Y_1,\ldots Y_{1078})$

A subsidiary experiment (simulation study, etc.) is used to estimate $A$. As described in \citep{lee:etal}, 
%the calibration scientists conducting 
the subsidiary measurements provide a sample of replicates of the vector $A$, where the variability in the sample is representative of the uncertainty in $A$. We denote this sample as ${\cal M} =\{A^{(1)},\ldots, A^{(M)}\}$, where $M$ is the number of replicates of $A$ in $\cal M$. The default statistical analysis uses a single default value of the vector $A_{\rm default}$, e.g., the mean of $\cal M$. We aim to account for the uncertainty in $A$ in our final estimates and error bars for $\alpha$ and $\beta.$
\citep{lee:etal} conducts a principal component analysis (PCA) of $\cal M$ to derive a lower dimensional variable, $Z$, that represents the bulk of %variance 
variation in $A$. Specifically, they derive $f$, where $A = f(Z)$ and $Z$ is a low-dimensional vector, of length 7 in our numerical results. PCA allows us to define $f$ in such a way that the components of $Z$ are each independent standard Gaussian variables, i.e. with mean 0 and variance 1. In our Bayesian analysis, we treat $Z$ (and hence $A=f(Z)$) as an unknown parameter with this Gaussian distribution as its prior distribution. 

\begin{figure}[p]
\begin{center}
\includegraphics[width=0.8\textwidth]{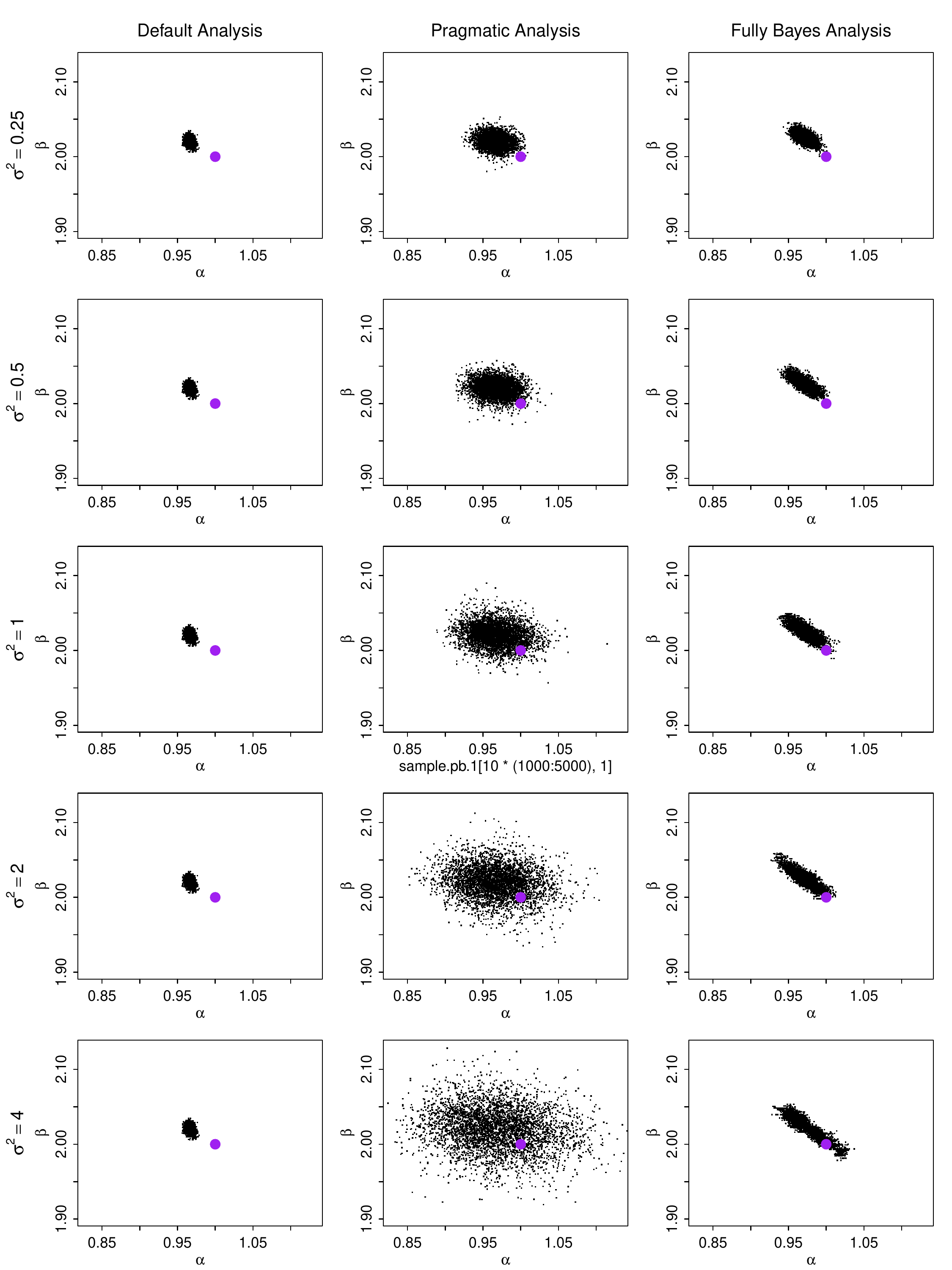}
\caption{Comparing the default, pragmatic, and fully Bayesian analysis in an example from astrophysics. 
The panels show MCMC samples from the posterior distribution of $\alpha$ and $\beta$, with the purple dot representing the true value. Columns correspond to the three analyses and rows compare different levels of uncertainty under the subsidiary analysis. The default analysis uses an estimate of the effective area from the subsidiary analysis, but ignores its uncertainty. This can lead to both bias and underestimation of uncertainty. The pragmatic analysis accounts for this uncertainty but does not use data from the primary experiment to refine estimation of the effective area. This can lead to overestimation of uncertainty when a significant proportion of information for the effective area stems from the primary experiment, as in the lower rows in this figure.}
\label{fig:calib-compare}
\end{center}
\end{figure}

We simulate a data set consisting of 1078 bin counts under the Poisson model with expectation given in Equation~\ref{dvd.model} using a particular effective area vector, $A_{\rm true}$, and with $\alpha = 1$ and $\beta = 2$. (These values of $\alpha$ and $\beta$ are marked with the large purple point in Figure~\ref{fig:calib-compare}).  $A_{\rm true}$ is selected to be consistent with the uncertainty in $A$ represented by $\cal M$, yet significantly offset from $A_{\rm default}$. 

We compare three Bayesian frameworks for estimating $\alpha$ and $\beta$, the parameters of scientific interest:
\begin{enumerate}
    \item {\it Default Analysis.} The default analysis fixes $A = A_{\rm default}$, ignoring uncertainty from the subsidiary experiment.  Thus, analysis is based on the conditional posterior distribution, $$\pi(\alpha, \beta \mid A= A_{\rm default}, Y).$$ Because the default analysis ignores uncertainty in $A$, its results quantify just the statistical uncertainties in $\alpha$ and $\beta$.
    \item {\it Pragmatic Analysis.} The pragmatic analysis accounts for uncertainty in $A$, but does not allow $Y$ to update this uncertainty. Thus, analysis is based on the distribution,
    $$\pi_{\rm prag}(\alpha, \beta, Z \mid Y) = \pi(\alpha, \beta \mid Z, Y) \pi(Z),$$ recalling that $A=f(Z)$ and where $\pi(Z)$ is the distribution that quantifies the uncertainty under the subsidiary analysis, i.e., normal distributions on the components of $Z$.
    \item {\it Fully Bayesian Analysis.} The fully Bayesian analysis uses all data to update all unknowns and is thus based on the posterior distribution,
    $$\pi_{\rm full} (\alpha, \beta, Z \mid Y) = \pi(\alpha, \beta \mid Z, Y) \pi(Z\mid Y),$$
    again recalling that $A=f(Z).$    
\end{enumerate}
In all three analyses, we use non-informative prior distributions on $\alpha$ and $\beta$. The pragmatic and fully Bayesian analyses set independent mean-zero Gaussian distributions on each of the seven components of $Z$. To explore how the level of uncertainty stemming from the subsidiary analysis combines with the statistical errors from the Poisson counts,
we consider five settings for the prior variances on $Z$. By construction, the (prior) variance of $Z$ is unity for the actual calibration of the Chandra X-ray observatory. This is the middle of the five settings that we consider, 
$\sigma^2 = 0.25, 0.5, 1, 2$ and $4$. 

Using Markov chain Monte Carlo (MCMC), we obtain a sample of size 4000 from the pragmatic Bayesian distribution and the fully Bayesian posterior distributions of $(\alpha, \beta, Z)$. This is repeated for each of the five settings of $\sigma^2$,
i.e., the levels of uncertainty in the subsidiary analysis. Using the default analysis, we similarly obtain a Markov chain Monte Carlo sample of size 4000 for $(\alpha, \beta)$. Because the default analysis ignores the uncertainty in the subsidiary analysis, it does not depend on $\sigma^2$ which quantifies this uncertainty, and thus need not be repeated. The results for $\alpha$ and $\beta$ appear in Figure~\ref{fig:calib-compare}, where rows correspond to five settings of $\sigma^2$ and columns correspond to the default, pragmatic, and fully Bayesian analyses. Because the default analysis does not depend on $\sigma^2$, it is thus the same in all five rows. 

Comparing the pragmatic and fully Bayesian results, if the subsidiary analysis is highly informative for $A$, the pragmatic and fully Bayesian methods give similar results in terms of the posterior means and variances of $\alpha$ and $\beta$. In this case the prior and posterior variances of the components of $Z$ under the fully Bayesian methods are similar.  As the subsidiary analysis becomes less informative (lower rows), however, the primary experiment provides most of the information for $Z$ and the pragmatic and fully Bayesian methods diverge in terms of the variance of the components of $Z$. This in turn means that the fully Bayesian method is able to better constrain $\alpha$ and $\beta$. Still the pragmatic Bayesian approach is clearly better than the default approach that severely underestimates the uncertainties in $\alpha$ and $\beta$. 

%Fig 4: David's plots
\vspace{0.1in}
It is interesting to compare the approaches used in this example and those used in Section~\ref{sec:line} for straight line fitting:

\begin{itemize}
\item{There are 2 parameters of interest in the Astronomy case ($\alpha$ and $\beta$), while the systematic effect is provided by the uncertainty in 
$A$. There is only one parameter of interest for the straight line (its gradient $b$), while its intercept $a$ is the nuisance parameter.} 
\item{The Astronomy example uses a Bayesian approach, while the straight line fit is based on likelihoods.}

\item{The straight line example uses the uncertainties in the main and the subsidiary experiments to obtain the expected uncertainty of the parameter of interest (the gradient), while the Astrophysics one uses a MCMC distribution to see the spread in the extracted parameters $\alpha$ and $\beta$.}
\end{itemize}

\section{Conclusions}

As mentioned in Sect. \ref{Other}, the Pragmatic approach can be simpler to apply in practice. 
%The Pragmatic method is usually simpler to apply in practice, especially for situations where the systematic is complicated. As an example from Particle Physics, most analyses using data from high energy accelerators  need detailed information about the behaviour of quarks and gluons within a proton; these are parametrised in the so-called `Parton Distribution Functions'. They are derived from global analyses of many other high energy processes, and also requires significant theoretical input; the uncertainties of such procedures are a source of systematic for the main analysis. The Pragmatic approach greatly simplifies the way in which these are incorporated in analyses.
However, it does not update the parameter(s) of interest as the $\nu$ are varied, and so can lose useful information. Furthermore, if the subsidiary experiment has a very large uncertainty, the estimate of the contribution to 
$\sigma _{\rm syst}$ can be unrealistically large. Also in its usual form, a single number is used to characterise the 
uncertainty on a nuisance parameter\footnote{Sometimes two numbers are used to give the upward and downward uncertainties
in $\nu$.}; the Full Likelihood is not restricted in this way. 

In contrast, the Full Likelihood method can become quite complicated when there are many subsidiary measurements 
providing information about the nuisance parameters. However, it benefits from the properties of likelihood methods,
in making efficient use of the data.  For the straight line example earlier, it produces results in line with intuition. The more realistic astrophysics example confirms this.

\vspace{0.1in}

The net conclusion is that the Full Likelihood is in general far preferable, and should be used when possible.

\vspace{0.1in}
ACKNOWLEDGEMENT

We wish to thank Tom Loredo for pointing us to the literature on modularization and cutting feedback. 

%We would like to thank..... %Bob Cousins, David van Dyk, Luc Demortier and Roberto Trotta for their advice on 
%various sections of this article

\vspace{0.3in}

\bigskip
\bibliography{systematics}

\end{document}